\documentclass{aa}  
\usepackage{graphicx}
\usepackage{txfonts}
\usepackage[colorlinks=true,linkcolor=blue,citecolor=blue,filecolor=blue,urlcolor=blue]{hyperref}
\defcitealias{K24}{KKV24}
\begin{document}

   \title{Effects of grain temperature distribution on organic protostellar envelope chemistry}
      \author{
		Juris Kalv\=ans	
    \and
		Juris Freimanis
		}
   \institute{
Engineering Research Institute `Ventspils International Radio Astronomy Centre' of Ventspils University of Applied Sciences,\\
In$\rm \check{z}$enieru 101, Ventspils, LV-3601, Latvia\\
\email{juris.kalvans@venta.lv}
			}
   \date{Received August 11, 2024; accepted November 18, 2024}
  \abstract
{Dust grains in circumstellar envelopes are likely to have a spread-out temperature distribution.}
{We aim to investigate how trends in temperature distribution between small and large grains affect the hot corino chemistry of complex organic molecules (COMs) and warm carbon-chain chemistry (WCCC).}
{A multi-grain multi-layer astrochemical code with an up-to-date treatment of surface chemistry was used with three grain temperature trends: grain temperature proportional to grain radius to the power -1/6 (Model M-1/6), to 0 (M0), and to 1/6 (M1/6). The cases of hot corino chemistry and WCCC were investigated, for a total of six models. The essence of these changes is for the main ice reservoir -- small grains -- having higher (M-1/6) or lower (M1/6) temperature than the surrounding gas.}
{The chemistry of COMs shows better agreement with observations in models M-1/6 and M1/6 than in Model M0. Model M-1/6 shows best agreement for WCCC because earlier mass-evaporation of methane ice from small grains induces the WCCC phenomenon at lower temperatures.}
{Models considering several grain populations with different temperatures can more precisely reproduce circumstellar chemistry.}
   \maketitle

\section{Introduction}
\label{intrd}

Chemical modelling of the interstellar medium, star-forming regions and protoplanetary disks has revealed importance of the dust temperature $T_d$ difference between small and large grains. For grains with size (radius) $a$ of about $\rm\mu$m or more, temperature decreases with larger $a$. For smaller grains, characteristic to the interstellar and circumstellar medium, studies indicate different trends. \citet{Li01,Cuppen06,Krugel07,Rollig13}, and \citet{Iqbal18} have found that generally the small grains with $a<$0.05\,$\rm \mu$m have the highest $T_d$, while \citet{Heese17,Sipila20}, and \citet{Gavino21} arrive at an opposite result, with large $\approx$0.2\,$\rm\mu$m grains having the highest $T_d$. The sub-$\rm\mu$m size range is dominant in a variety of astrophysical environments, such as diffuse medium, photodissociation regions, molecular clouds, and protostellar envelopes.

We aim to analyse the astrochemical significance of the grain-size temperature dependence problem for star-forming regions with a multi-layer astrochemical model. Instead of attempting to produce a version of $T_d$ distribution between sub-$\rm\mu$m grains, we simply investigate two main cases, first, smaller grains having a higher temperature than larger grains $T_{d,\rm small}>T_{d,\rm big}$ and, second, $T_{d,\rm small}<T_{d,\rm big}$. An intermediate case with $T_{d,\rm small}=T_{d,\rm big}$ was also considered. In a positive scenario, comparison of calculation results with published observations of gas and ice molecules can indicate an astrochemically preferential trend in $T_d$ distribution between grain sizes.

A practical goal of this study was the application and adaptation of the recently improved multi-grain multi-layer model \textsc{Alchemic-Venta} for describing evaporating ices, a necessary capability for further research. As far as we know, astrochemical models considering multiple grain types have not yet been applied for the case of evaporating ices in protostellar envelopes.

\section{Methods}
\label{mthd}

In our recent work \citep[][hereinafter KKV24]{K24} we developed a rate-equation astrochemical model with a detailed description of gas-grain chemistry. It combines multi-sized grains and multi-layered ices, where the desorption (or surface binding) energy $E_D$ of icy molecules partially depends on their surrounding environment, be it bare grains, polar icy species, or non-polar ices. In addition to evaporation and surface diffusion, other important $E_D$-dependent processes added were desorption by the exothermic H+H surface reaction, photodesorption, and chemical desorption. It turned out that multiple sorts of $E_D$-dependent desorption mechanisms shape the formation of interstellar ices. The gas-grain chemical balance is changed by a rapid chemical desorption and a fast adsorption due to a higher surface area with the multi-grain approach.

The multi-grain multi-layer approach of \citetalias{K24} focusses on microscopic surface processes and thus the model is well adapted for the aims of this study. A short explanation of the model's features is provided below. Because the model is the same as in \citetalias{K24}, we keep the description concise and focus on the few changes introduced for the purposes of this study.

\subsection{Chemical model}
\label{mchm}

\begin{table}
\caption{Desorption-related updates to surface species: list of species added to the H-bond rule with $E_D$ reduced by 500\,K on completely bare grains.}
\label{tab-Hb}
\centering
\tiny
\begin{tabular}{l l l l}
\hline\hline
\multicolumn{4}{c}{Additions to the H-bond rule} \\
\hline
H$_2$CCO & CH$_3$COCH$_3$ & H$_2$CO & HC$_7$N \\
CH$_3$CHO & CH$_3$O & HC$_3$N & HC$_9$N \\
CH$_3$CN & H$_2$CN & HC$_5$N &  \\
\hline
\end{tabular}
\end{table}

We employ the UMIST \textsc{Rate2012} gas-phase network of \citet{McElroy13} combined with the OSU surface chemistry network of \citet{Garrod08}. The latter has been reduced to account only for species included in \textsc{Rate2012}. The initial chemical abundances are listed in Table~1 of \citetalias{K24} and have been adapted from \citet{Wakelam08} and \citet{Garrod13}.

Neutral gas-phase species are accreted onto the grains, where they are sequentially arranged in three bulk-ice layers and a surface layer. Surface and bulk-ice species are subjected to photodissociation (modified by a factor of 0.3, relative to gas phase species) and binary reactions. For the latter, the ratio between $E_D$ and surface diffusion energy $E_{\rm diff}$ was taken to be 0.5.  The activation energy barriers $E_A$ for the CO$_2$-generating surface reactions $\rm CO+O$ and $\rm CO+OH$ important for ice chemistry, are 630\,K and 176\,K, respectively.

The value of $E_D$ for each molecule varies, depending on the surroundings (bare grain, H$_2$O- or CO-dominated ices) in each layer and each grain size bin \citepalias[Sects. 2.4 and 2.5 of][]{K24}. On bare grain carbonaceous surfaces, hydrogen bonds are unable to form, thus the molecular $E_D$ is reduced for according to the `H-bond rule' described with Table~5 of \citetalias{K24}. Here we extended the H-bond rule for strongly polar species with a dipole moment in excess of 2\,D that do not have a H atom attached to an electronegative atom or functional group, listed in Table~\ref{tab-Hb}. The value of $E_D$ on completely bare grains is reduced by 500\,K for these species, compared to $E_D$ on water surface. Such a change reflects the ability of these species to contribute an electron pair for a H-bond in a watery environment and the inability to do so on a carbonaceous or silicate surface.

The core is irradiated by interstellar photons (Habing field), attenuated by interstellar extinction $A_V$, and cosmic rays (CRs), attenuated by the column of matter between the margin of the cloud and its centre. The hydrogen atom column density $N_H=2.2\times10^{21}A_V$\,cm$^{-2}$.

The model considers six mechanisms for surface molecule desorption: thermal evaporation, photodesorption by interstellar and CR-induced photons, CR-induced whole-grain heating, chemical desorption of products in exothermic surface reactions, and indirect chemical desorption by H+H combination reaction on the grains. Following the conclusions of \citet{Sipila19} and \citetalias{K24}, chemical desorption efficiency $f_{cd}$ from icy surfaces (i.e., excluding bare grains) was reduced by a factor of 0.1 for reactions producing hydrogen nitrides.

\subsection{Temperature model}
\label{tgr}

\begin{table}
\caption{Models with different temperature $T_d$ dependence on grain size and physical conditions in the prestellar stage.}
\label{tab-Td}
\centering
\tiny
\begin{tabular}{l c c c}
\hline\hline
Model\tablefootmark{a} & $y$ in Eq.~(\ref{tgr1}) & $T_{d,0.037}$\tablefootmark{b},\,K & $T_{d,0.232}$\tablefootmark{b},\,K \\
\hline
M-1/6 & -1/6 & 118 & 87 \\
M0 & 0 & 100 & 100 \\
M1/6 & 1/6 & 85\tablefootmark{c} & 115 \\
\hline\hline
Model\tablefootmark{d} & $n_{\rm fin}$,\,cm$^{-3}$ & $T_{\rm max}$,\,K & $f_{\rm CR}$ \\
\hline
COMs & $10^8$ & 200 & 0.1 \\
WCCC & $10^6$ & 50 & 1.0 \\
\hline
\end{tabular} \\
\tablefoottext{a}{See Sect.~\ref{tgr}.}
\tablefoottext{b}{Example $T_d$ for smallest ($T_{d,0.037}$) and largest ($T_{d,0.232}$) grains, when $T_{\rm gas}$=100\,K.}
\tablefoottext{c}{$T_{d,0.037}$ can be higher by 2...6\,K because of $\approx50$\,\% increase in grain radius due to the icy mantle.}
\tablefoottext{d}{See Sect.~\ref{mcr}.}
\end{table}
%

\begin{figure}
		\hspace{-2cm}
 \includegraphics{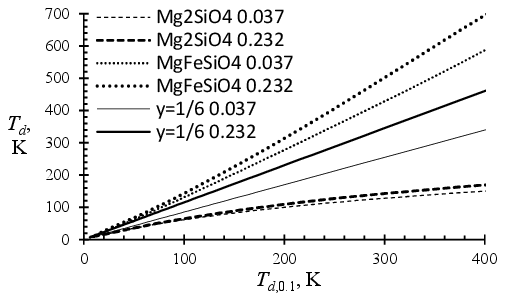}
		\vspace{-23cm}
 \caption{Grain temperature difference for four types of silicate grains as a function of their average temperature $T_{d,0.1}$. Numbers in the legend indicate size in $\rm\mu$m. For comparison, the only-size-dependent $T_d$ employed by our model are also shown, with $y$=1/6. (The case with $y$=-1/6 is almost identical, with the large- and small-grain lines swapping places.) The value of $T_{d,0.1}$ on the x-axis practically depends on the amount of radiation received (or, therefore, distance) from the protostar.}
 \label{fig-Td}
\end{figure}

Five grain size bins were considered covering a size range of 0.03 to 0.3\,$\rm\mu$m. The bins are logarithmically distributed with the MRN interstellar grain size distribution. The radius of the smallest grains is $a=0.037\,\rm\mu$m and that of the largest grains is 0.232\,$\rm\mu$m (Table~3 of \citetalias{K24}).

In the prestellar stage, gas temperature $T_{\rm gas}$ and $T_d$ were calculated as functions of $A_V$ based on the star-forming region simulation results by \citet{Hocuk16} and \citet{Hocuk17}, respectively. The latter study provides temperature $T_{d,0.1}$ for a `classical' grain radius of 0.1\,$\rm\mu$m. This value was then used to obtain $T_d$ for grain sizes actually used in the simulations.

As discussed in Sect.~\ref{intrd}, several complex temperature calculations for small and large grains show opposite trends. For our aims, we adopt the simple grain size-temperature dependence of \citet{Krugel07}, where $T_d\propto a^y$, where the power $0\leq y < 1$
\begin{equation}
	\label{tgr1}
T_{d,a} = \left(\frac{a+b}{0.1}\right)^y \times T_{d,0.1}\,.
\end{equation}
Here, $a$ is grain radius ($\rm\mu$m) and $b$ is ice thickness ($\rm\mu$m), with \textit{a+b} being the full radius of a grain. For a trend, where the small grains have higher $T_d$, we take $y$=-1/6 \citep[Model `M-1/6',][]{Krugel07,Pauly16}. For the opposite trend, we simply employ $T_d\propto a^{1/6}$ ($y$=1/6, Model `M1/6'). \citet{Li01} and \citet{Heese17} indicate a less-steep $T_d$ dependence on grain radius and likely $-1/6<y<1/6$. Therefore, with Model `M0' we consider a case where all grains have equal $T_d$. These models are further divided in two versions adapted for different groups of organic compounds (Sect.~\ref{mcr}). The model information is summarised in Table~\ref{tab-Td}.

Positive $y$ is supported by a simple test calculation of dust temperature around a low-mass with protostar, made for this study with the RADMC-3D program\footnote{https://www.ita.uni-heidelberg.de/~dullemond/software/radmc-3d/} \citep[e.g.,][]{Commercon12}. The protostar was assumed to have a 4700\,K surface temperature and luminosity of 3\,L$_\odot$. Grain temperature varies with distance from the star (i.e., amount of electromagnetic radiation energy received) and grain size $a$. Fig.~\ref{fig-Td} shows that an even more important factor, than size is grain material, of which two types of amorphous interstellar silicate were considered -- MgFeSiO$_4$ and Mg$_2$SiO$_4$. With this method, $y<0.1$, or a difference $T_{d,0.232}-T_{d,0.037}$ of $<$10\,K between the largest 0.232\,$\mu$m and smallest 0.037\,$\mu$m grains when $T_{d,0.1}\approx100$\,K. If we differentiate grain populations by their materials, not sizes, the $T_d$ difference for grains of similar sizes and different materials, $T_{d,\rm MgFeSiO4}-T_{d,\rm Mg2SiO4}$, is significantly greater at $\approx$70\,K. Such a result is a strong indication that notable $T_d$ differences between circumstellar grain populations do exist in the same time and place. Here we stick to $T_d$ dependence on $a$, continuing the work in \citetalias{K24}, although implementing $T_d$ dependence on grain materials can be relatively straightforward.

The temperatures in the protostellar stage were anchored at $T_{d,0.1}$. As described in Sect.~\ref{mcr}, $T_{d,0.1}$ rises from $<$10\,K at the end of the prestellar stage to 200\,K at the protostellar envelope stage. The $T_d$ for the five grain size bins were calculated from $T_{d,0.1}$ with the help of Eq.~\ref{tgr1}, except for Model M0, where $T_d=T_{d,0.1}$ for all bins.

Gas temperature changes in relation to $T_d$ in circumstellar environments \citep{Chapillon08,Podolak11,Gavino21}. For a study focussing on the chemical effects of different $T_d$ trends in five models, for $T_{\rm gas}$ we need simple approach that does not complicate the interpretation of gas-grain chemistry. Therefore, we assumed that $T_{\rm gas}=T_{d,0.1}$ in all models. This approach is in line with the above studies showing that $T_{\rm gas}$ is similar to the $T_d$ of sub-micron grains.

\subsection{Macrophysical model}
\label{mcr}

\begin{figure}
		\hspace{-2cm}
 \includegraphics{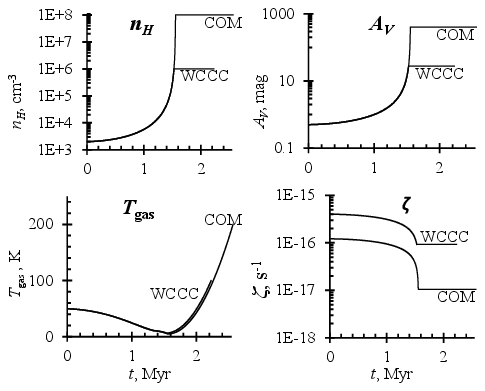}
		\vspace{-22cm}
 \caption{Evolution of hydrogen atom numerical density $n_H$, interstellar extinction $A_V$, gas temperature $T_{\rm gas}$, and the cosmic-ray ionisation rate $\zeta$ in the COM and WCCC models.}
 \label{fig-phys}
\end{figure}

The model applies a standard approach for modelling an interstellar molecular star-forming cloud core. A parcel of matter near the centre of a low-mass 2\,M$_\odot$ core is considered with initial $A_V$=0.5\,mag and hydrogen atom number density $n_H$=2000\,cm$^{-3}$. The core undergoes a free-fall collapse to a final density $n_{\rm fin}$, depending on model (see below). We assume that at this point the protostar has formed and density increase stops. The core has transformed into a protostellar envelope and the gas parcel heats up according to the $T_2$ time-dependence of \citet[][i.e., $T_{d,0.1}\propto t_{\rm pr}^2$, where $t_{\rm pr}$ is time after birth of the protostar]{Garrod06}, reaching $T_{\rm gas}$ of 200\,K over 1\,Myr after the formation of the protostar. While \citet{Garrod06} included also shorter heating time-scales, only the long 1\,Myr time-scale was considered here in order to allow discerning the gas-mediated movement of molecules between the five grain size bins. Evaporation from multiple grain types with shorter heating time-scales is being investigated in another recently submitted study.

The above conditions are relevant for hot corinos with abundant COMs \citep{Cazaux03} and warm carbon-chain chemistry (WCCC) cores \citep{Sakai08}, which allows us to compare the simulation results of these compounds to their observed abundances. Both groups of species are affected by the distribution trend of grain temperatures (expressed by the variable $y$) in the protostellar envelope stage. These effects become apparent mostly when the gas temperature rises relatively slowly, when evaporated molecules can re-freeze again onto colder grains in a different grain size bin. This process, along with the uneven distribution of ice mass between grain size bins, is what produces differences between models with different $T_d$ trends. Therefore, we employ a long time-scale of 1\,Myr for the heat-up. In effect, the model describes the isochoric heat-up of a gas parcel approaching the protostar from the depths of the envelope.

Hot corinos and WCCC cores differ by their physical conditions. The former have $n_H\approx10^7...10^9$\,cm$^{-3}$ and `hot' temperatures of a few hundred K. The latter are characterized by lower densities of $n_H\approx10^6$\,cm$^{-3}$ and `lukewarm' conditions with $T_{\rm gas}$ of only several tens of K. Furthermore, WCCC is apparently caused by external irradiation of the star-forming core \citep{Lindberg15,Taniguchi21oph}, with cosmic rays being the most important ionisation agent \citep{K21,Sun24}. Many cores are between these two extremes, showing both C-chains and COMs \citep[e.g.,][]{Mercimek22}.

Taking the above into account, the three models M-1/6, M0, and M1/6 are divided in two versions each, one for the COM chemistry, and one for WCCC. The COMs models have $n_{\rm fin}=10^8$\,cm$^{-3}$ and a maximum final gas temperature $T_{\rm max}$ of 200\,K, while for WCCC models $n_{\rm fin}=10^6$\,cm$^{-3}$ and $T_{\rm max}=100$\,K (Table~\ref{tab-Td}). Consequently, the WCCC models reach $n_{\rm fin}$ at integration time $t$=1.53\,Myr and the protostellar stage ends at 2.23\,Myr. For the COMs models, these times are 1.55 and 2.57\,Myr, respectively.

Regarding irradiation, the cosmic-ray ionisation rate $\zeta$ was calculated as a function of $N_H$ following \citet{Ivlev15p}, model `High'. For the COMs model, this value of $\zeta$ was modified by a factor $f_{\rm CR}$=0.1, while for the WCCC model $f_{\rm CR}$=0.3. This approach was borrowed from \citet{K21}, where it was found that CRs are the primary agent promoting WCCC activity, while decreasing the abundance of COMs. Other CR phenomena, such as CR-induced photon flux or whole-grain heating were modified proportionally to $\zeta$. The value of $\zeta$ varies from $1.2\times10^{-16}$ to $1.0\times10^{-17}$\,s$^{-1}$ for the COMs model and from $4.0\times10^{-16}$ to $9.3\times10^{-17}$\,s$^{-1}$ for the WCCC. Higher $\zeta$ for the latter model can be caused for example, by exposure to unattenuated CR flux at the margins of molecular clouds or by nearby massive star-forming regions \citep[see the discussion and references in][]{K21}. Fig.~\ref{fig-phys} shows the basic physical conditions in the COM and WCCC models.

\section{Results}
\label{rez}

\subsection{General chemistry}
\label{gnrl}

\begin{figure}
		\hspace{-2cm}
 \includegraphics{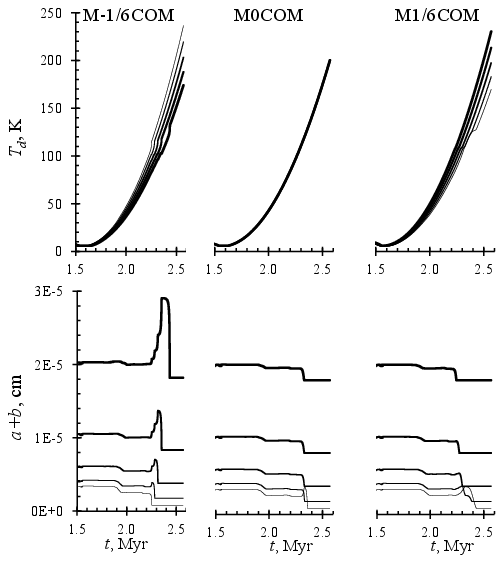}
		\vspace{-17cm}
 \caption{Dust temperature and size evolution in the protostellar stage. Thicker lines are for larger grains. The kinks in $T_d$ curves above 100\,K are due to rapid grain size change with H$_2$O ice evaporation.}
 \label{fig-gr}
\end{figure}
%

\begin{figure}
	\vspace{-2cm}
\includegraphics{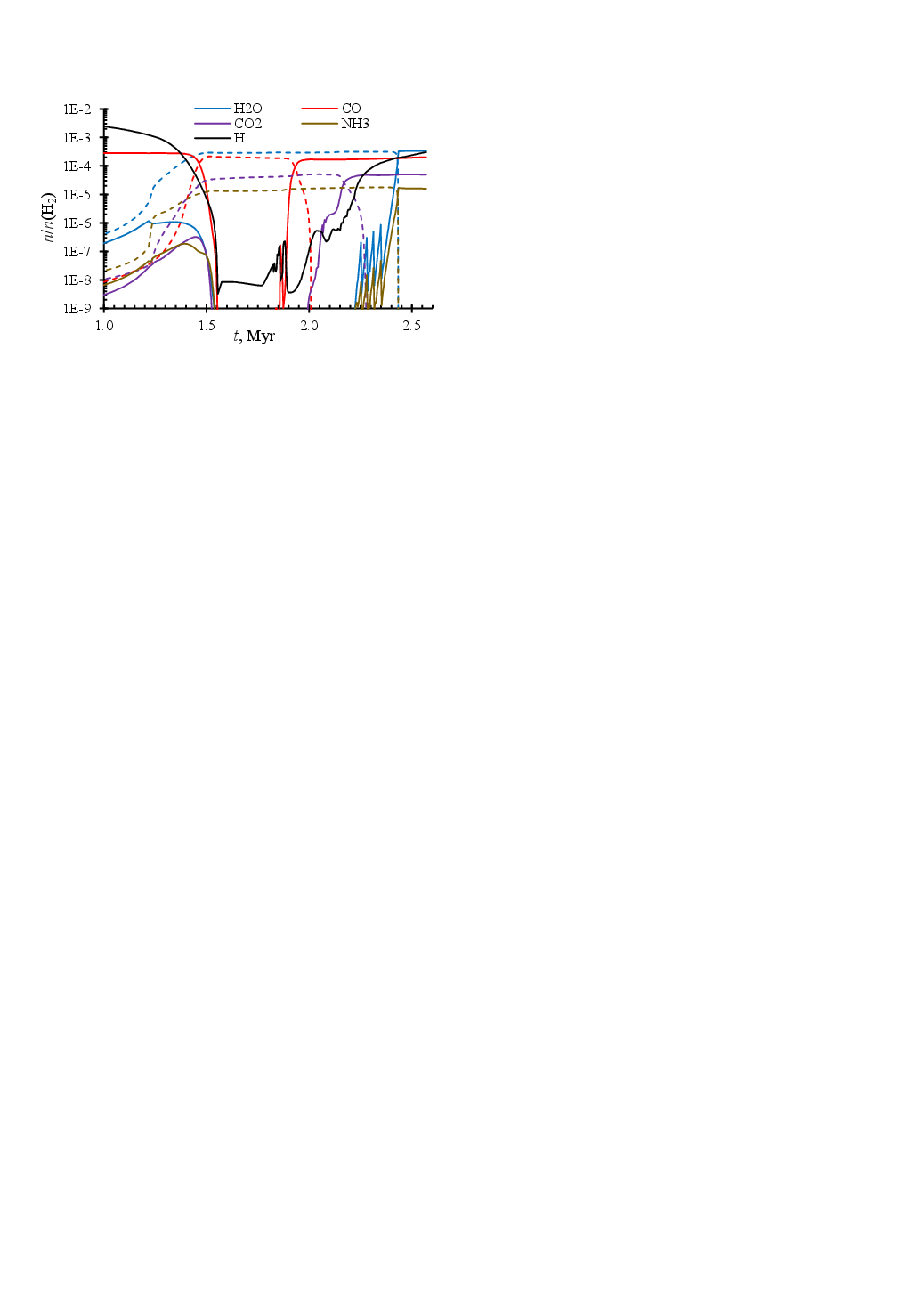}
	\vspace{-19cm}
	\caption{Overall chemical results for the freeze-out period and the protostellar stage in Model M-1/6COM: abundances of selected major species in gas (solid lines) and ices (dashed lines).}
	\label{fig-gen}
\end{figure}

The general results of this model for the prestellar core stage depend little on the $T_d$ distribution and are discussed in detail by \citetalias{K24}. Fig~\ref{fig-gr} shows the evolution of grain temperature and size (\textit{a+b}) for the five models in the protostellar envelope stage. Grains enter the protostellar stage with ice thickness $b$ in the range of 66...82 monolayers (MLs), with higher $b$ for smaller grains in all models. The abundant smallest 0.037$\rm\mu$m grains carry about 42\,\% of all icy molecules at the end of the prestellar stage. The main molecular process during the heat-up of the envelope is evaporation of volatiles from warmer grains accompanied by partial conversion into less-volatile species and their freeze-out over long ($>$0.1\,Myr) time-scales and followed by their immediate partial re-freeze onto colder grains (in M-1/6 and M1/6) over shorter times (a few 10$^4$\,yr). For the long heating time-scales and high gas density considered in the COM models, the re-freeze can be nearly complete.

Overall gas-grain chemistry differs little between the five models. Model M-1/6COM is most similar to \citetalias{K24} and we use it for describing the overall chemical results. Fig.~\ref{fig-gen} shows the evolution of abundance, relative to H$_2$, for gaseous and solid major species in the protostellar and prestellar stages with Model M-1/6COM. A characteristic feature for this model is a see-saw pattern with four gas-phase abundance peaks for H$_2$O and species evaporating together with H$_2$O ice owing to successive evaporation and re-freeze from four grain size bins when they reach $T_d>100$\,K. Molecules evaporated from the fifth grain size bin have no colder grains where to freeze-out again and their transition to the gas phase is permanent. The saw is less pronounced in Model M1/6COM with positive $y$ (i.e., higher $T_d$ for larger grains), because most of ice mass is on the small grains, which evaporate last, while naturally there no such peaks in Model M0COM with equal $T_d$ for all grains.

Increase of grain size due to re-freeze in Model M-1/6COM lowers $T_d$ for grains currently experiencing ice accumulation and makes the repeated freeze-out more efficient (see also Fig.~\ref{fig-coms}). Although the gas abundance of H$_2$O varies by three orders of magnitude in the see-saw period (2.20--2.38\,Myr), it actually represents less than 0.3\,\% of the overall H$_2$O abundance. The amount of H$_2$O ice ($3.1\times10^{-4}$ relative to H$_2$), integrated over all grain sizes, practically does not change, until H$_2$O evaporation from the hottest small grains commences at 2.38\,Myr and $T_{d, 0.037}$=108\,K in Model M-1/6COM.

Interestingly, to some extent, the re-freeze process is present even in Model M0COM, as molecules evaporated from larger grains freeze onto smaller grains, although they all share equal $T_d$. Such a phenomenon is possible because until $T_d<120$\,K the accretion rate onto the 0.037\,$\rm \mu$m grains is faster than thermal desorption, thanks to the large surface area of the finely dispersed small grains. The accretion rate is boosted by evaporation from other grain size bins that increases gas-phase abundance of H$_2$O and other species with similar $E_D$, temporarily shifting the accretion-evaporation balance towards freeze-out onto small grains.

\subsection{COMs in the hot corino}
\label{karsc}

\begin{figure*}
	\centering
	\hspace{-2cm}
\includegraphics{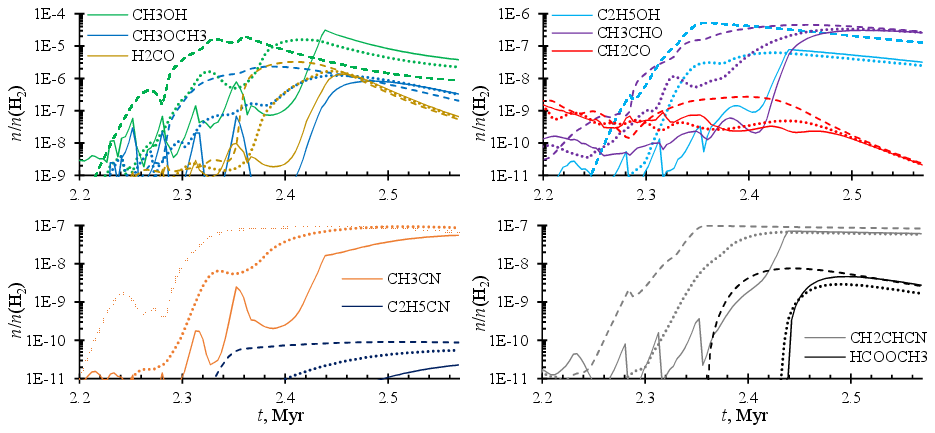}
	\vspace{-21cm}
	\caption{Calculated abundances of COMs in the late protostellar stage with $T_{\rm gas}\geq90$\,K. Solid lines are for Model M-1/6COM, dashed lines for M0COM, and dotted lines for M1/6COM.}
	\label{fig-coms}
\end{figure*}
%

\begin{table*}
\caption{Comparison of observed \citep{Drozd19,Jorgensen20} and calculated average (150--200\,K) and final (200\,K) gas-phase abundances, relative to methanol for COMs and other species.}
\label{tab-coms}
\centering
\tiny
\begin{tabular}{ l | c | c c c | c c c }
\hline\hline
 & Observations & \multicolumn{3}{c}{Model average 150--200\,K} & \multicolumn{3}{c}{Model end point 200\,K} \\
Species & IRAS16293B & M-1/6COM & M0COM & M1/6COM & M-1/6COM & M0COM & M1/6COM \\
\hline
H$_2$O & 3.3E+2 & 3.0E+1 & \textbf{1.9E+2}\tablefootmark{a} & \textbf{5.9E+1} & \textbf{8.9E+1} & \textbf{3.9E+2} & \textbf{1.5E+2} \\
CO & 1.0E+1 & \textbf{1.8E+1} & 1.1E+2 & \textbf{3.3E+1} & \textbf{5.4E+1} & 2.3E+2 & \textbf{8.9E+1} \\
CH$_3$OH & \textit{1.0E+19}\tablefootmark{b} & \textit{5.4E-6}\tablefootmark{c} & \textit{8.8E-7}\tablefootmark{c} & \textit{2.9E-6}\tablefootmark{c} & \textit{1.9E-6}\tablefootmark{c} & \textit{4.3E-7}\tablefootmark{c} & \textit{1.1E-6}\tablefootmark{c} \\
H$_2$CO & 1.9E-1 & \textbf{5.0E-2} & \textbf{4.2E-1} & \textbf{1.2E-1} & 1.7E-2 & \textbf{6.2E-2} & \textbf{2.8E-2} \\
C$_2$H$_5$OH & 2.3E-2 & \textbf{4.8E-3} & \textbf{1.2E-1} & \textbf{7.2E-3} & \textbf{8.4E-3} & \textbf{1.5E-1} & \textbf{1.1E-2} \\
CH$_3$OCH$_3$ & 2.4E-2 & \textbf{5.2E-2} & 4.3E-1 & \textbf{1.5E-1} & \textbf{8.7E-2} & \textbf{2.3E-1} & \textbf{1.4E-1} \\
HCOOCH$_3$ & 2.6E-2 & 3.0E-4 & \textbf{3.0E-3} & 3.6E-4 & 7.3E-4 & \textbf{3.0E-3} & 7.5E-4 \\
CH$_3$CHO & 1.2E-2 & \textbf{2.2E-2} & 2.2E-1 & \textbf{5.1E-2} & \textbf{6.8E-2} & 3.3E-1 & \textbf{1.2E-1} \\
HCOOH & 5.6E-3 & \textbf{4.7E-3} & 7.4E-2 & \textbf{1.7E-2} & \textbf{1.5E-2} & 1.5E-1 & \textbf{4.5E-2} \\
CH$_3$COCH$_3$ & 1.7E-3 & 1.8E-6 & 6.8E-5 & 4.5E-6 & 7.2E-6 & 6.7E-5 & 1.1E-5 \\
HCN & 5.0E-3 & 8.0E-2 & 1.3E+0 & 3.4E-1 & 4.2E-1 & 3.4E+0 & 1.2E+0 \\
CH$_3$CN & 4.0E-3 & \textbf{3.3E-3} & 4.8E-2 & \textbf{1.5E-2} & \textbf{1.5E-2} & 7.7E-2 & \textbf{3.9E-2} \\
C$_2$H$_5$CN & 3.6E-4 & 1.0E-6 & \textbf{5.0E-5} & 7.0E-6 & 6.0E-6 & \textbf{1.0E-4} & 2.5E-5 \\
CH$_2$CHCN & 7.4E-5 & 6.0E-3 & 5.0E-2 & 1.1E-2 & 1.6E-2 & 9.6E-2 & 2.6E-2 \\
CH$_2$CO & 4.8E-3 & 1.3E-5 & 2.4E-4 & 3.6E-5 & 5.6E-6 & 2.7E-5 & 9.6E-6 \\
HNCO & 3.7E-3 & 8.8E-2 & 4.9E-1 & 1.9E-1 & 2.3E-1 & 8.7E-1 & 4.1E-1 \\
HC$_3$N & 1.8E-5 & 1.8E-2 & 1.4E-1 & 4.2E-2 & 6.3E-2 & 3.5E-1 & 1.3E-1 \\
H$_2$S & 1.7E-2 & 2.6E-10 & 4.5E-11 & 1.2E-11 & 6.3E-13 & 2.1E-12 & 9.6E-13 \\
OCS & 2.5E-2 & 2.0E-4 & \textbf{6.8E-3} & 2.1E-3 & 2.8E-5 & 3.4E-4 & 1.3E-4 \\
CS & 3.9E-4 & 1.6E-5 & \textbf{2.9E-4} & \textbf{9.1E-5} & \textbf{9.5E-5} & \textbf{1.2E-3} & \textbf{4.6E-4} \\
H$_2$CS & 1.3E-4 & 4.5E-3 & 8.4E-2 & 2.6E-2 & 1.4E-2 & 1.8E-1 & 7.1E-2 \\
SO$_2$ & 1.3E-4 & 9.8E-3 & 4.3E-6 & \textbf{2.9E-4} & 2.8E-2 & 1.1E-5 & \textbf{7.2E-4} \\
SO & 4.4E-5 & \textbf{1.5E-4} & 1.9E-7 & 3.8E-6 & 7.7E-4 & 5.6E-7 & \textbf{1.9E-5} \\
\hline
$\bar{\kappa}$ & . . . & \textbf{0.321} & 0.276 & 0.313 & 0.247 & 0.234 & 0.286 \\
$\bar{\kappa}_{\rm red}$ & . . . & \textbf{0.463} & \textbf{0.403} & \textbf{0.474} & \textbf{0.374} & \textbf{0.353} & \textbf{0.442} \\
Fits &  & 8 & 7 & 10 & 8 & 7 & 11 \\
\hline
\end{tabular}
\\
\tablefoottext{a}{Calculated values agreeing within a factor of 10 with observed values are marked with bold font.}
\tablefoottext{a}{Column density, cm$^{-2}$.}
\tablefoottext{b}{Column density, cm$^{-2}$.}
\tablefoottext{c}{Abundance relative to H$_2$, $n{\rm(CH_3OH)}/n{\rm(H_2)}$.}
\end{table*}

Heavy molecules in protostellar envelopes arise in thermal desorption from ice-covered grains, with limited processing in the gas. As a means for sampling circumstellar organic synthesis in ices, gaseous COMs have raised significant interest. More than a dozen COMs are routinely detected in low-mass Class 0/I star-forming cores \citep{Ceccarelli23}.

Thermal desorption from grains of a typical icy COM starts with its desorption from the outer surface of the hottest grain size bin, experiences re-freeze onto colder grains, until the next grain size heats up sufficiently. Figure~\ref{fig-coms} shows that many of the COM gas-phase abundance curves show the four-peak saw from the evaporation -- re-freeze cycles in Model M-1/6COM. The highest abundance achieved in these cycles is at least an order of magnitude lower than a species' abundance after its full evaporation.

These cycles end either with a species' diffusion out of the bulk-ice of the coldest remaining grain size bin that has been heated too much to sustain these species in its ice layer or the species' evaporation and non-thermal desorption from a near-bare dust surface. The latter scenario occurs after the evaporation of major species H$_2$O, NH$_3$, and CH$_3$OH, when no bulk-ice remains for the COMs to hide in. Most COMs have their desorption or bulk-ice diffusion energies comparable to those of water. Decrease of H$_2$O ice abundance by a factor of 10 occurs first for Model M0COM at $t$=2.35\,Myr, gradually from 2.35 to 2.42\,Myr for M1/6COM, and at 2.43\,Myr for M-1/6COM. This sequence generally is obeyed also by COMs. It is regulated by the inability for grains to sustain H$_2$O ice at $T_d>108$\,K, although evaporation of major remaining ice components begins already at 102\,K. For Model M-1/6COM, the final ice evaporation event occurs from the low number of grains in the largest size bin with a nearly 400\,ML thick ice layer. The accompanying increase of grain size delays the programmed increase of large-grain temperature $T_{d,\rm 0.0232}$, even temporarily reversing it at $T_{d,\rm 0.0232}\approx103$\,K (2.35\,Myr). This effect causes an unexpected delay in ice evaporation for Model M-1/6COM, compared to M0COM and M1/6COM, where no such dramatic change of grain size occurs. H$_2$O ice evaporation ends at $T_d\approx$125\,K, predictably a lower temperature than that obtained from temperature-programmed laboratory experiments with astrophysical ices \citep{Fayolle11,Martin14}.

Table~\ref{tab-coms} compares the calculated gas-phase COM abundances, relative to CH$_3$OH for Models M-1/6COM, M0COM, and M1/6COM, along with observational data for the low-mass protostar IRAS 16293B surveyed with ALMA. To make maximum use of modelled data, we applied a twofold comparison with the with observations: average abundances in the 150--200\,K interval, when all ices have evaporated ($t$=2.43...2.57\,Myr), and the final abundances at 200\,K. Average abundances are useful here because the time-evolution of the model, tracking an infalling parcel of matter, qualitatively represents the 1D chemical structure of the protostellar envelope, effectively crudely mimicking the observational line-of-sight \citep{Garrod06}. We adopted a difference within an order of magnitude as the basic measure for an `agreement' -- a fit between modelled and observed abundances. The final row of Table~\ref{tab-coms} shows that the number of such fits is highest for Model M1/6COM for both, average and final abundance comparison. To evaluate this result statistically, we applied the `confidence' parameter $\kappa_i$ of \citet[][see also \citeauthor{Song24} \citeyear{Song24}]{Garrod07}
\begin{equation}
	\label{res1}
\kappa_i = erfc \left(\frac{|{\rm lg}(X_i)-{\rm lg}(X_{{\rm obs},i})|}{\sqrt{2}\sigma}\right)\,,
\end{equation}
where $X_i$ is the calculated (relative) abundance of species $i$, $X_{{\rm obs},i}$ is the observed abundance, and $\sigma$ was assumed to be unity. For evaluating modelling results, we used the $\bar\kappa$ value averaged over all entries in Table~\ref{tab-coms} and $\bar\kappa_{\rm red}$ for a reduced number of species, namely those that show observational agreement at least for one of the models. In effect, $\bar\kappa_{\rm red}$ excludes observed molecules with possible contamination from regions with significantly different physical conditions and molecules with insufficient chemical network in the model. Table~\ref{tab-coms} shows that the $\kappa$ parameters are lowest for Model M0COM in all cases, while it is generally highest for Model M1/6COM. If one draws a single average value from the four $\bar\kappa$ values available for each model, then the value for such a `$\bar\kappa_\kappa$' parameter is 0.351, 0.316, and 0.379 for Models M-1/6COM, M0COM, and M1/6COM, respectively.

Model M0COM displays a lower agreement with observations, implying that gradual evaporation of ices from grains with a spread-out $T_d$ distribution better explain the observations. The early evaporation of icy COMs (primarily, CH$_3$OH) in Model M0COM causes their over-processing in the gas, producing an overabundance of CH$_3$OH derivatives, such as H$_2$CO, C$_2$H$_5$OH, CH$_3$OCH$_3$, and CH$_3$CHO. The spread in $T_d$ in Models M1/6COM and M-1/6COM mitigates the over-processing, especially by denying high CH$_3$OH abundances at gas temperatures of 125--145\,K (2.35--2.41\,Myr), when CH$_3$OH is transformed into H$_2$CO with a high efficiency. Model M1/6COM shows the highest number of fits with observed abundances and the highest $\bar\kappa_\kappa$, and is thus the best-agreeing model for the hot-corino chemistry.

\subsection{WCCC}
\label{silcc}

\begin{figure*}
	\hspace{-2cm}
\includegraphics{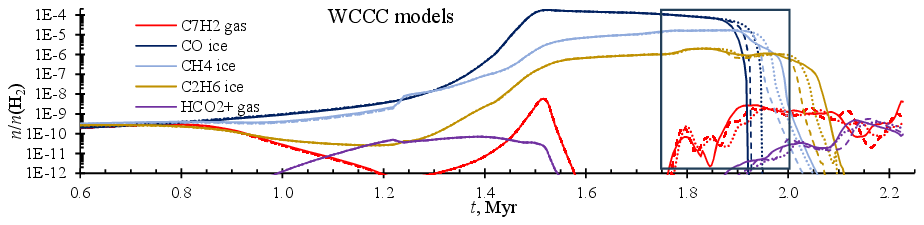}
	\vspace{-24.5cm}
	\caption{Correlation of an example WCCC molecule C$_7$H$_2$ with evaporation of C-containing ices CO, CH$_4$, and C$_2$H$_6$. Solid lines are for Model M-1/6WCCC, dashed lines for M0WCCC, and dotted lines for M1/6WCCC. The box at $t$=1.8...2.1\,Myr ($T_{\rm gas}$=20...70\,K) outlines the window with C-chain abundance peaks, most relevant for WCCC with this model.}
	\label{fig-wccc}
\end{figure*}

\begin{figure*}
	\hspace{-2cm}
\includegraphics{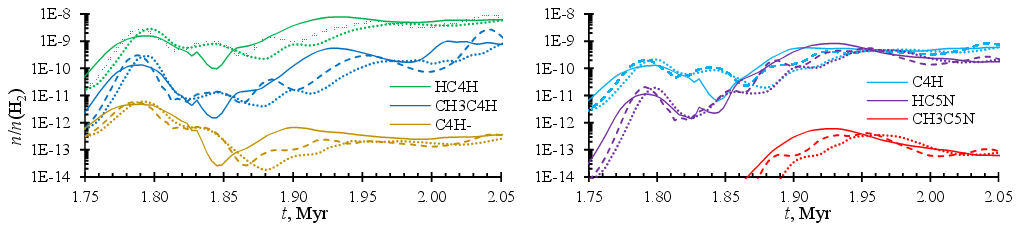}
	\vspace{-24.5cm}
	\caption{Example calculated carbon-chain abundances in the WCCC window. Solid lines are for Model M-1/6WCCC, dashed for M0WCCC, and dotted for M1/6WCCC.}
	\label{fig-chains}
\end{figure*}

Carbon chains can be described with a general chemical formula RC$_n$R, where the functional groups R can be either non-existent or H, CH$_3$, CN, S, or other atoms. Fig.~\ref{fig-wccc} shows the evolution of the gas-phase abundance of an example carbon chain, C$_7$H$_2$. Its abundance initially peaks at the prestellar stage at $t\approx$1.1\,Myr, when $A_V\approx1.2$\,mag and $n_H\approx10^4$\,cm$^{-3}$. The next peak occurs near the star-formation time, when density has rapidly increased and irradiation level dropped, allowing different molecules to form. These peaks occur in a starless cloud and are produced by cold interstellar chemistry.

In the protostellar envelope, C$_7$H$_2$ reappears with an uneven abundance curve. Periods with increase of abundance are primarily caused by evaporation of hydrocarbon ices. The evaporated species are dissociated in the gas by CR-induced photons, creating the WCCC phenomenon. It produces longer chains that promptly freeze-out, again locking away carbon atoms and ending the current gas-phase C-chain abundance bump, until temperature rises sufficiently for the next cycle \citep{Taniguchi19}. Because of a lower $\zeta$, the bumps are spread-out, compared to those of \citet{K21}. The newly-frozen chains reside on the outer surface of the grains, which means that CR-induced ice photoprocessing still maintains elevated gas-phase C-chain abundances between the bumps.

Unlike the hot corino model in the above section, multiple peaks associated with species' successive evaporation from the five grain size bins were not obtained for carbon chains. This was because of lower (a few K) $T_d$ differences between the bins and a slow rise in temperature. Evaporation from different grain size bins and ice layers effectively overlaps, while lower $n_H$ means a lower rate for gas molecule re-accretion onto grain surfaces. Instead, COMs evaporate at conditions of $T_d$ differences of tens of K, when temperatures are rapidly increasing, causing sudden evaporation of volatiles, while the higher $n_H$ allows for a rapid re-freeze process, producing the four abundance peaks in Model M-1/6COM above $T_{\rm gas}$=100\,K.

Detected observational conditions for WCCC are characterized by low beam-averaged temperatures of 10--15\,K \citep{Sakai08,Sakai09lupus} and evaporation of methane ice \citep{Sakai12}. Thus, we attribute to WCCC only the first few C-chain abundance peaks, occurring in the $t$ interval of 1.8--2.0\,Myr and 20--50\,K $T_{\rm gas}$ for all WCCC models. Elevated C-chain abundances continue up to 2.1\,Myr and gas temperatures of 70\,K, which we deem too high to represent the WCCC phenomenon. Fig.~\ref{fig-wccc} shows that the C-chain chemistry is initially reactivated in the protostellar stage after the evaporation of CO ice. This event is accompanied with a partial re-freeze of carbon and oxygen in the form of CH$_4$ and H$_2$O ices. Non-thermal desorption of outer-surface CH$_4$ ice then starts low-level C-chain activity in the gas. The next major peak occurs with the thermal desorption of CH$_4$ ice, starting at $T_d>$34\,K, and continues with the evaporation of C$_2$H$_6$ ice at $T_d>$45\,K. The peaks occur at he same $t$ and $T_{\rm gas}$ for all C-chains in a given simulation.

Fig.~\ref{fig-chains} shows example WCCC abundances for chains with the general formula RC$_4$R. The C-chain abundance peaks shift between simulations with different grain temperature distributions. The peak due to methane evaporation occurs at $t\approx$1.92; 1.95, and 1.97\,Myr for Models M-1/6WCCC, M0WCCC, and M1/6WCCC, respectively. This sequence can be explained with the temperature of the primary ice reservoir (small grain size bins, $T_{d,0.037}$) relative to that of gas. When $T_{d,0.037}>T_{\rm gas}$ (M-1/6), evaporation of hydrocarbon ices occurs earlier and WCCC is observed at lower $T_{\rm gas}$, and vice versa. The peak due to evaporation of ethane ice is visible only for some species at 2.00--2.04\,Myr.

Table~\ref{tab-wccc} in the Appendix compares the calculated C-chain abundances with observed values. The prototypical WCCC core L1527 has the best assortment of astrochemical observations and was used as a reference. To quantify agreement with observations, we compared average abundance in the 20--50\,K with observed values, again assuming that evolution in time partially represents the 1D chemical make-up of the protostellar envelope. This comparison yields indistinguishable results for all models because the average abundances usually differ within a factor of few. The values of $\bar\kappa$ and $\bar\kappa_{\rm red}$ are close to 0.47 and 0.59 in all cases, respectively. The number of calculated $n/n_{\rm H_2}$ falling within a factor of 10 from observations is close to 20.  A similar conclusion was reached when we used maximum abundances instead of the average ones.

In order to further determine differences between the M-1/6, M0, and M1/6 WCCC model results, we compare gas temperature $T_{\max}$ of a species' highest abundance peak in 15--50\,K interval. Observations indicate that the WCCC phenomenon occurs at low temperatures of $>$13\,K \citep{Sakai13}, while the C-chain abundance peaks in our models are closer to 40\,K. Therefore, we presume that lower $T_{\rm max}$ values indicate that a model is better at reproducing WCCC. Evaluation by $T_{\rm max}$ clearly favours Model M-1/6WCCC because the C-chain peaks have a temperature sequence M-1/6WCCC -- M0WCC -- M1/6WCCC, meaning that for M-1/6WCCC the peaks consistently occur at the lowest $T_{\rm gas}$. The latter aspect is due to the primary ice reservoir (small grains) having a higher temperature than $T_{\rm gas}$ in M-1/6WCCC and evaporating their ices earlier, at lower $T_{\rm gas}$. This results lends more support to a negative $y$ with small grains having higher temperatures than large grains. It is shown quantitatively with Table~\ref{tab-wccc}, where peak abundances from a minimum $T_{\rm gas}$ of 15\,K (at $t$=1.74\,Myr) of the protostellar stage up to 50\,K ($t$=2.01\,Myr) are given. We assume that a significant $T_{\rm max}$ difference between the models is 3\,K because it produces a 2 or more magnitude difference in the evaporation rate of CH$_4$ and C$_2$H$_6$ ices. Comparing to our previous study on WCCC \citep{K21}, the C-chain abundance maxima have been shifted to higher $T_{\rm gas}$ because of a higher adopted $E_D$ for CH$_4$ and thus higher evaporation temperature for CH$_4$ ices.

\section{Conclusions}
\label{sechi}

The comparison of calculated abundances of COMs with those observed in hot corinos shows that Models M-1/6COM and M1/6COM give better agreement for relative proportions between the various COMs than Model M0COM. Regarding carbon chains, Model M-1/6WCCC has its C-chain gas-phase abundance peaks at lower gas temperature, facilitating the association with the WCCC phenomenon from observations. This result is also supported by the observation of the HCO$_2^+$ ion at a WCCC source \citep[see also Fig.~\ref{fig-wccc}]{Sakai09lupus}. The abundance of this ion rises with the desorption of CO$_2$ ice, which also occurs at lower $T_{\rm gas}$ in Model M-1/6WCCC. In this respect, Model M1/6WCCC shows the least agreement with these observational trends, while M0WCCC is in the middle. The comparison of observed and calculated abundances yields similar results for all three WCCC models.

The chemically important difference between all models is that in M-1/6 the main ice reservoir (small grains) has a higher temperature than that of gas, in M1/6 a minority of ices have $T_d>T_{\rm gas}$ and majority have $T_d<T_{\rm gas}$, while for M0 all ices always have $T_d=T_{\rm gas}$. The latter is physically least feasible because, given the potential diversity of dust, all circumstellar grains cannot be expected to have a similar $T_d$. Their temperatures differ not only because of different sizes but even more so because of their chemical composition (Fig.~\ref{fig-Td} in Sect.~\ref{tgr}). The M0 models serve as `in situ' comparison of our calculation results with simpler, single-grain-type models, commonly used in astrochemistry. We will consider grain populations made up from different materials in a future study.

Concluding, this investigation indicates that the introduction of the multi-grain approach clearly affects circumstellar organic chemistry. For the interstellar cloud -- prestellar collapse stage, on the contrary, Models M0, M-1/6, and M1/6 produce chemical results that are similar for most species. Notably, the calculated composition of interstellar ice was found to be practically equal, underlining that the changes in protostellar organic chemistry are produced solely by differing ice evaporation patterns in the heated protostellar envelope model.

\begin{acknowledgements}
This research is funded by the Latvian Science Council grant `Desorption of icy molecules in the interstellar medium (DIMD)', project No. lzp-2021/1-0076 and has made use of NASA’s Astrophysics Data System. We thank the referee for significant improvements to the presentation.
\end{acknowledgements}

   \bibliographystyle{aa}
   \bibliography{protostar}

\begin{appendix}
\section{Results of the WCCC models}

\begin{table*}
\caption{Calculated average and maximum abundances and gas temperature $T_{\rm max}$ during the WCCC window of 20--50\,K (1.81--2.01\,Myr) for carbon-chains and other species', relative to H$_2$.}
\label{tab-wccc}
\tiny
\begin{tabular}{ l | c c | c c c | c c c |c c c }
\hline\hline
 & \multicolumn{2}{c}{Observations} & \multicolumn{9}{c}{Model} \\
 & \multicolumn{2}{c}{L1527} & \multicolumn{3}{c}{M-1/6WCCC} & \multicolumn{3}{c}{M0WCCC} & \multicolumn{3}{c}{M1/6WCCC} \\
Molecule & $n/n(\rm H_2)$ & Ref.\tablefootmark{a} & $n/n(\rm H_2,max)$ & $T_{\rm gas}$, K & $n/n(\rm H_2,avg)$ & $n/n(\rm H_2,max)$ & $T_{\rm gas}$, K & $n/n(\rm H_2,avg)$ & $n/n(\rm H_2,max)$ & $T_{\rm gas}$, K & $n/n(\rm H_2,avg)$ \\
\hline
CO & 3.9E-5 & 1 & 1.8E-5 & 34 & \textbf{8.8E-6}\tablefootmark{b} & 1.4E-5 & 36 & \textbf{6.6E-6} & 1.3E-5 & 39 & \textbf{5.0E-6} \\
NH$_3$ & 1.9E-8 & 2 & 7.6E-8 & 27 & \textbf{3.8E-8} & 7.4E-8 & 30 & \textbf{3.7E-8} & 7.4E-8 & 31 & \textbf{3.8E-8} \\
CN & 2.4E-9 & 3 & 1.1E-8 & 36 & \textbf{5.7E-9} & 9.5E-9 & 40 & \textbf{4.7E-9} & 8.9E-9 & 42 & \textbf{4.7E-9} \\
HCN & 1.2E-9 & 3 & 2.0E-8 & 37 & \textbf{9.3E-9} & 2.5E-8 & 40 & \textbf{9.6E-9} & 2.4E-8 & 42 & \textbf{1.0E-8} \\
HNC & 3.2E-10 & 3 & 6.9E-9 & 37 & 3.9E-9 & 8.5E-9 & 40 & 4.0E-9 & 8.4E-9 & 41 & 4.4E-9 \\
HCO$^+$ & 6.0E-10 & 3 & 4.5E-9 & 50 & \textbf{2.5E-9} & 4.5E-9 & 50 & \textbf{2.1E-9} & 4.2E-9 & 50 & \textbf{1.8E-9} \\
N2H$^+$ & 2.5E-10 & 3 & 2.8E-9 & 27 & \textbf{9.5E-10} & 3.0E-9 & 30 & \textbf{1.0E-9} & 3.1E-9 & 32 & \textbf{1.2E-9} \\
SO & 1.4E-10 & 3 & 5.9E-12 & 25 & 2.1E-12 & 7.6E-12 & 27 & 1.7E-12 & 5.7E-12 & 28 & 1.7E-12 \\
CS & 3.3E-10 & 3 & 5.4E-11 & 30 & 1.9E-11 & 6.3E-11 & 27 & 2.1E-11 & 5.8E-11 & 28 & 2.1E-11 \\
\hline
CH$_4$ & 9.8E-6 & 4 & 3.5E-6 & 37 & \textbf{1.0E-6} & 3.9E-6 & 40 & 8.9E-7 & 2.8E-6 & 42 & 7.1E-7 \\
CH$_3$CN & 9.1E-12 & 5 & 4.1E-10 & 37 & 1.6E-10 & 5.5E-10 & 40 & 1.6E-10 & 5.5E-10 & 42 & 1.8E-10 \\
CH$_2$CN & 7.9E-11 & 5 & 4.6E-11 & 35 & \textbf{1.9E-11} & 3.3E-11 & 31 & \textbf{1.1E-11} & 2.5E-11 & 33 & \textbf{9.9E-12} \\
CH$_3$OH & 8.2E-10 & 5 & 1.5E-10 & 30 & 3.5E-11 & 3.1E-11 & 43 & 2.0E-11 & 3.3E-11 & 34 & 1.7E-11 \\
 & 2.3E-9 & 6 &  &  &  &  &  &  &  &  &  \\
\hline
C$_2$H & $>$1.8E-8 & 7 & 1.1E-8 & 35 & 5.9E-9 & 1.2E-8 & 19 & 4.8E-9 & 1.2E-8 & 20 & 4.4E-9 \\
HC$_3$N\tablefootmark{d} & 9.6E-10 & 6 & 8.4E-9 & 38 & \textbf{2.9E-9} & 7.6E-9 & 40 & \textbf{2.1E-9} & 6.2E-9 & 42 & \textbf{1.9E-9} \\
 & 4.4E-10 & 5 &  &  &  &  &  &  &  &  &  \\
CH$_3$CCH & 2.1E-9 & 7 & 2.4E-9 & \textbf{38}\tablefootmark{c} & \textbf{1.0E-9} & 2.2E-9 & 41 & \textbf{8.2E-10} & 1.9E-9 & 44 & \textbf{7.1E-10} \\
C$_2$S & 1.8E-10 & 7 & 2.2E-11 & 24 & 5.4E-12 & 3.9E-11 & 26 & 8.4E-12 & 2.8E-11 & 26 & 7.8E-12 \\
CH$_2$CO & 2.5E-10 & 5 & 2.7E-10 & 50 & \textbf{5.4E-11} & 2.3E-10 & 50 & \textbf{3.0E-11} & 1.1E-10 & 50 & 1.9E-11 \\
\hline
C$_3$H & 2.0E-10 & 8 & 4.0E-9 & \textbf{38} & \textbf{1.6E-9} & 3.8E-9 & 41 & \textbf{1.2E-9} & 3.7E-9 & 44 & \textbf{1.3E-9} \\
 & 1.2E-10 & 6 &  &  &  &  &  &  &  &  &  \\
c-C$_3$H$_2$ & 4.6E-10 & 6 & 9.6E-9 & 50 & 4.6E-9 & 8.5E-9 & 50 & \textbf{3.2E-9} & 7.4E-9 & 47 & \textbf{2.7E-9} \\
l-C$_3$H$_2$ & 3.9E-11 & 7 &  &  &  &  &  &  &  &  &  \\
\hline
C$_4$H\tablefootmark{d} & 6.8E-9 & 7 & 2.9E-10 & \textbf{34} & 1.5E-10 & 2.3E-10 & 41 & 1.1E-10 & 2.4E-10 & 45 & 1.0E-10 \\
 & 3.5E-8 & 9 &  &  &  &  &  &  &  &  &  \\
C$_4$H$^-$ & 3.9E-13 & 10 & 2.4E-12 & 20 & \textbf{4.0E-13} & 2.8E-12 & 19 & \textbf{2.8E-13} & 3.0E-12 & 20 & \textbf{3.6E-13} \\
C$_4$H$_2$ & 5.7E-11 & 7 & 3.9E-9 & \textbf{38} & 1.7E-9 & 2.6E-9 & 41 & 1.1E-9 & 2.2E-9 & 45 & 9.1E-10 \\
HC$_5$N & 1.9E-10 & 7 & 4.1E-10 & \textbf{38} & \textbf{1.3E-10} & 2.8E-10 & 41 & \textbf{8.5E-11} & 2.3E-10 & 42 & \textbf{7.5E-11} \\
 & 8.6E-11 & 5 &  &  &  &  &  &  &  &  &  \\
CH$_3$C$_4$H & 2.0E-10 & 8 & 4.2E-10 & 50 & \textbf{9.4E-11} & 1.5E-10 & 19 & \textbf{4.4E-11} & 1.5E-10 & 20 & 3.9E-11 \\
\hline
C$_5$H & 3.2E-11 & 7 & 1.7E-10 & \textbf{38} & \textbf{7.0E-11} & 1.3E-10 & 41 & \textbf{4.8E-11} & 1.2E-10 & 44 & \textbf{4.3E-11} \\
 & 1.9E-11 & 5 &  &  &  &  &  &  &  &  &  \\
\hline
C$_6$H & 2.2E-11 & 11 & 7.9E-11 & 34 & \textbf{3.4E-11} & 6.2E-11 & 36 & \textbf{2.4E-11} & 6.1E-11 & 44 & 2.3E-11 \\
 & 3.9E-11 & 8 &  &  &  &  &  &  &  &  &  \\
C$_6$H$_2$ & 3.9E-11 & 11 & 2.1E-9 & \textbf{38} & 8.5E-10 & 1.5E-9 & 41 & 5.4E-10 & 1.3E-9 & 44 & 4.7E-10 \\
 & 6.6E-12 & 8 &  &  &  &  &  &  &  &  &  \\
HC$_7$N & 5.4E-11 & 7 & 2.5E-10 & \textbf{38} & \textbf{7.0E-11} & 1.6E-10 & 41 & \textbf{3.8E-11} & 1.3E-10 & 44 & 3.3E-11 \\
\hline
C$_7$H & 2.1E-12 & 8 & 2.2E-11 & 34 & \textbf{8.7E-12} & 1.9E-11 & 36 & \textbf{6.3E-12} & 1.8E-11 & 40 & 6.0E-12 \\
HC$_9$N & 5.0E-12 & 7 & 1.0E-10 & \textbf{38} & \textbf{2.9E-11} & 6.4E-11 & 41 & \textbf{1.6E-11} & 5.4E-11 & 42 & 1.4E-11 \\
\hline
$\bar\kappa$ &  &  &  &  & \textbf{0.468} &  &  & \textbf{0.467} &  &  & \textbf{0.462} \\
$\bar\kappa_{\rm red}$ &  &  &  &  & \textbf{0.599} &  &  & \textbf{0.593} &  &  & \textbf{0.585} \\
Fits &  &  &  & 9 & 20 &  & 0 & 20 &  & 1 & 19 \\
\hline
\end{tabular} \\
\tablefoottext{a}{1 -- \citet{Jorgensen02}, 2 -- \citet{Hirota09}, 3 -- \citet{Jorgensen04}, 4 -- \citet{Sakai12}, 5 -- \citet{Yoshida19}, 6 -- \citet{Sakai09deut}, 7 -- \citet{Sakai08}, 8 -- \citet{Araki17}, 9 -- \citet{Sakai10}, 10 -- \citet{Sakai08-}, 11 -- \citet{Sakai07}.  Where necessary, $N_{\rm H_2}$ was taken to be $2.8\times10^{22}$ \citep{Jorgensen02}.}
\tablefoottext{b}{Calculated relative abundances agreeing within a factor of 10 with observed values are marked with bold font.}
\tablefoottext{c}{$T_{\rm max}$ values lower by at least 3\,K than those of any of the other models are marked with bold font.}
\tablefoottext{d}{More than an order of magnitude lower abundances have been observed in the WCCC core Chamaeleon MMS1 \citep{Cordiner12cha}.}
\end{table*}

\end{appendix}

\end{document}